\documentclass[compsoc, conference, a4paper, 10pt, times]{IEEEtran}

\usepackage{cite}
\usepackage{amsmath,amssymb,amsfonts}
\usepackage{algorithmic}
\usepackage{graphicx}
\usepackage{textcomp}
\usepackage{xcolor}
\usepackage{booktabs}
\usepackage[hidelinks]{hyperref}
\usepackage{listings}
\usepackage{subcaption}
\usepackage{pgfplots}
\usepgfplotslibrary{groupplots}
\pgfplotsset{compat=newest}
\usepackage{makecell}
\begin{document}
	
\lstdefinelanguage{ASM}{
	keywords=[1]{
		b, ble, blt, bne, ldr, str, mov, f32f32, nop, add, shl, stib, untyped, ld, u32, 1d, imm, rpt2, nop3, rpt5, s2ur, umov, uldc, uimad, stg, e, sys, exit, wide, 64, texture\_sample, wait, fadd32, mov\_imm, writeout, st\_tile, stop, get\_sr, device\_store},
	keywords=[2]{
		w, x, y, z, r0, r51, r2, c8, R1, UR6, SR\_CTAID, UR7, UR4, UR7, R8, sp, r1l, r1h, r2l, ts0, ss0, r3, r1l\_r1h\_r2l, r0\_r1\_r2\_r3, discard, u2\_u3, thread\_pos\_in\_grid, sr80 },
	sensitive=false,
	morestring=[b]',
	morecomment=[l]{//},
	basicstyle=\footnotesize\ttfamily,
	keywordstyle=[1]\bfseries\color{black},
	keywordstyle=[2]\color{green!40!black},
	identifierstyle=\color{black},
	commentstyle=\itshape\color{purple},
	stringstyle=\color{orange},
}

\lstdefinelanguage{algorithm} {
	mathescape=true,
	frame=tB,
	numbers=left, 
	numberstyle=\tiny,
	basicstyle=\scriptsize, 
	keywordstyle=\color{black}\bfseries\em,
	keywords={,input, output, return, datatype, function, in, if, for, else, foreach, while, begin, end, }
	numbers=left,
	xleftmargin=.04\textwidth,
	morecomment=[l]{//},
	commentstyle=\itshape\color{purple}
}

\title{Whispering Pixels: Exploiting Uninitialized Register Accesses in Modern GPUs}

% Submissions should be anonymized. See the CFP for details on how to anonymize your paper, including any references to your own work.
%\author{\em Anonymous Authors}

% The author information is skipped here, but can be used to include author information in the publication.
\iftrue
\author{\IEEEauthorblockN{Frederik Dermot Pustelnik, Xhani Marvin Sass, Jean-Pierre Seifert}
\IEEEauthorblockA{\textit{Technische Universität Berlin - SecT} \\
Berlin, Germany \\
\{f.pustelnik, sass, jean-pierre.seifert\}@tu-berlin.de}
}
\fi

\iffalse
\author{\IEEEauthorblockN{Anonymous Authors}
	\IEEEauthorblockA{\textit{} \\ \\
		\ }
}
\fi

\maketitle

\begin{abstract}
Graphic Processing Units (GPUs) have transcended their traditional use-case of rendering graphics and nowadays also serve as a powerful platform for accelerating ubiquitous, non-graphical rendering tasks. One prominent task is inference of neural networks, which process vast amounts of personal data, such as audio, text or images. Thus, GPUs became integral components for handling vast amounts of potentially confidential data, which has awakened the interest of security researchers. This lead to the discovery of various vulnerabilities in GPUs in recent years.\\
In this paper, we uncover yet another vulnerability class in GPUs: We found that some GPU implementations lack proper register initialization routines before shader execution, leading to unintended register content leakage of previously executed shader kernels. We showcase the existence of the aforementioned vulnerability on products of 3 major vendors - Apple, NVIDIA and Qualcomm. The vulnerability poses unique challenges to an adversary due to opaque scheduling and register remapping algorithms present in the GPU firmware, complicating the reconstruction of leaked data. In order to illustrate the real-world impact of this flaw, we showcase how these challenges can be solved for attacking various workloads on the GPU. First, we showcase how uninitialized registers leak arbitrary pixel data processed by fragment shaders. We further implement information leakage attacks on intermediate data of Convolutional Neural Networks (CNNs) and present the attack's capability to leak and reconstruct the output of Large Language Models (LLMs).

\end{abstract}

% Depending on how vigilant their paper processor is, IEEE may ask for these in your final paper, but we've heard about these amazing inventions called search engines that are able to index every word in your paper, so no need to include them in your submission unless you really want to.

%\begin{IEEEkeywords}
%component, formatting, style, styling, insert
%\end{IEEEkeywords}

\section{Introduction}
\label{sec:intro}

In recent years, the ubiqity of Graphic Processing Units (GPUs) has brought unprecedented computational power into the hands of end-users. Growing beyond their traditional use cases of rendering graphics only, GPUs have become a popular choice for parallelizing heavy workloads. Advancements in GPU architectures also fueled advancements in the field of artificial intelligence (AI), where the availability of increased processing power lead also to enhanced AI models. Typically, AI is used for the evaluation of large datasets, from processing medical data to automated evaluation of surveilance feeds. Today, AI suitable for everyday use has arrived at the hands of the end-user in the form of Large Language Models (LLMs) \cite{vaswani2017attention, touvron2023llama}. The recent popularity of such LLMs is primarily thanks the ability to provide meaningful answers and the ability to reason, a skill which has been mostly absent in previous research on natural language processing (NLP). Hence, a lot of business cases for the utilization of LLMs have emerged, ranging from automatically handling customer service requests to tasks such as text content generation. Datacenter providers have also noticed this increasing trend of AI advancements and offer GPU-as-a-Service (GaaS), where users can rent GPUs on demand and pay via a usage model. Customers can enjoy a high degree of flexibility, renting GPU computational power on demand. Providers also aim to reduce underutilized resources as much as possible. Allocating a complete GPU to a single container or virtual machine (VM) can be considered a waste of resources if either only a fraction or only a limited time of processing power is needed. The ability to share GPUs gives the provider a cost-effective measure to adapt to varying workloads. As industrial examples, Google's Kubernetis Engine allows to share a GPU between up to 48 tenants, while Microsoft build GPU Paravirtualization into the Hyper-V Hypervisor, allowing VMs to share a single GPU \cite{gketimeshare, hypervgpu}. \\
In a landscape also defined by connectivity and mobility, neural networks have also found their way into customer's end-devices. Smartphone GPU vendors offer the ability to leverage on-board graphics for acceleration of computational tasks. Real-time object recognition, facial tracking, video feed enhancement, audio classification and even voice cloning are just some examples on how users interact with neural networks on a daily basis.  \\
Vulnerabilities within GPUs can thus lead to adversaries gaining unauthorized access to private user data.

The primary feature of GPUs on end-user devices is rendering graphical user interfaces, which display potentially confidential information. To protect against information leakage attacks, including pixel information leakage, modern operating systems deploy tight sandboxing and access management in order to enforce hard boundaries between applications. Thus, GPUs are components which process confidential user data in different ways, which moreover has attracted the interest of security researchers. Researchers have presented multiple vulnerabilities in modern GPUs to bypass application boundary mechanisms, but mostly evaluated their attacks for specifically leaking pixel data by applying filters on webbrowser iframes \cite{taneja2023hot, wang2023gpu}. Other recently presented GPU side-channel attacks range from frequency-based attacks to leakage via uninitialized memory \cite{di2013cuda}. Naghibijouybari et al. presented a performance counter attack in order to leak typed letters on the on-screen keyboard \cite{naghibijouybari2018rendered}. Cache side-channels have also been replicated on GPUs \cite{jiang2016complete, jiang2017novel}. \\

In this paper, we present multiple register-based vulnerabilities found across different GPU implementations. For one, we demonstrate how uninitialized registers can be abused in order to gain access to computational results of previously executed shader kernels. Furthermore, we briefly discuss pixel leakage attacks and thoroughly evaluate the found vulnerabilities on state-of-the art neural networks in order to highlight the criticality of the presented flaws. Our attack displays a much higher data leakage count than other, similar previously discovered leakage attacks.

Our main contributions are:
\begin{itemize}
    \item We describe uninitialized GPU register access as a new side-channel technique which affects multiple vendors.
    \item We evaluate our attack for multiple data exfiltration attacks against different types neural networks, such image-processing neural networks and LLMs. Further, we showcase the attack's capability for leaking pixels across application boundaries.
    \item We showcase how uninitialized structures can be used to build a covert channel, which is, to the best of our knowledge, the fastest GPU-based covert channel.
    \item We propose multiple countermeasures against the aforementioned attacks and weigh the advantages and downsides of each approach.
\end{itemize}
\textbf{Outline} We structure the paper as follows: First, we describe the fundamentals of GPU architectures and the GPU pipeline in section \ref{sec:background}. Next, in section \ref{sec:threatmodel} we provide the considered threat model. Afterwards, in section \ref{sec:attack}, we describe the attacks and give insight into the building blocks required for successful exploitation. In section \ref{sec:evaluation}, we evaluate the performance and the exploitability of our attack for various different scenarios. Next, in section \ref{sec:countermeasures}, we propose multiple countermeasures and further discuss limitations and other aspects of our work in section \ref{sec:discussion}. Related work is presented in section \ref{sec:relatedwork}. Lastly, we conclude our paper in section \ref{sec:conclusion}.    % basic introduction
\section{Background}
\label{sec:background}
In this section, we furnish insight into the GPU pipeline, the differences and similarities of different GPU architectures and General Purpose Computation for GPUs, which is the main building block for executing non-graphical tasks, such as accelerated artificial intelligence applications.

\subsection{GPU Architectures}
\begin{figure}
    \centering
    \includegraphics[width=0.4\textwidth]{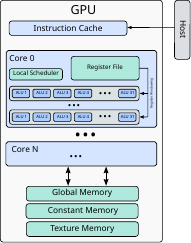}
    \caption{Generalized GPU architecture. A GPU typically contains one or more cores, where each core can contain multiple execution units, which can execute e.g. 32 threads as a wavefront in lockstep.}
    \label{fig:gpu-architecture}
\end{figure}
Many different GPU architectures by different vendors exist, however, they all share some similarities. In this section, we provide an overview over the generalized architecture as depicted in Figure \ref{fig:gpu-architecture} and will provide an overview over differences of various vendor implementations later. \\
\textbf{Threading Model}: The base of all GPUs is formed by a \textit{Core}, representing an execution unit which contains one or multiple SIMD units which in turn consist of multiple ALUs. Dispatched work is divided into multiple chunks which are executed in parallel on a SIMD unit. \\
\textbf{Registers}: All architectures have a set of vector registers, which are accessible for general-purpose usage in the shader. \\
\textbf{GPU Programming}: An application has to use vendor-provided libraries in order to translate a shader from its high-level source code to an architecture-dependent binary code. Vendors provide these libraries for a variety of high-level languages. As such, most vendors build their efforts on the LLVM compiler framework \cite{nvcc, amdllvm}, due to its extensibility and flexibility. This holds for open-source solutions, as the one provided by AMD, but also closed-source solutions, i.e. by NVIDIA and Qualcomm. To the application, the same programming interface is exposed (sometimes with some vendor-specific functions), such that shader source code can be compiled to any architecture with ease. In the rest of this paper, we will refer to the high-level source code, such as the OpenGL Shading Language (GLSL), as \textit{shader source code} and the architecture-dependent compilation result as \textit{shader binary}. The executed shader can also be referred as \textit{shader kernel}. \\
\textbf{Memory Hierarchy}: Each core has access to a dedicated cache, as well as a shared cache. GPUs further include dedicated memories for different kind of data, such as Texture Memory, Constant Memory and Global Memory for sharing data between kernels.\\
In order to support the parallel execution of multiple threads on a SIMD units, memory accesses have to be optimized. As a solution, GPUs support memory coalescing, where multiple memory accesses are bundled into a single large one, drastically improving throughput. Coalescing is one of the main reasons why vendors state that branching should be avoided in shader code.

\subsection{GPU Rendering Pipeline}
The simplified GPU rendering pipeline consists of three main stages: \\
\textbf{Geometry}: The geometry stage performs the transformation from the world coordinate system to screen coordinates for vertices. Perspective and clipping handling are also included in this step. Generally, the geometry stage is handled via a \textit{vertex shader}. \\
\textbf{Rasterization}: During rasterization, the GPU determines which pixels correspond to a given primitive. In many rendering pipelines, the rasterization process cannot be determined by the programmer via a shader. Based on the rasterization information, the next pipeline stage is only run on pixels belonging to a specific primitive. \\
\textbf{Fragmentation}: In this pipeline stage, the color, shading and texturing of all primitives are determined. Each fragment, which is associated with one pixel on the screen, is the result of a \textit{fragment shader} calculation. A fragment contains all data to calculate the pixel value, which may include the depth, interpolated color or texture values and the pixel position. \\
With the continuous advancements of graphic rendering techniques, such as ray-tracing, the rendering pipeline outgrew the traditional designs and is thus more complex for advanced 3D renderers. These steps are not important to understand the rest of the paper and hence have been omitted.

\subsection{General Purpose Computation on GPU (GPGPU)}
\begin{table*}[!h]
\begin{center}
\caption{Vendor-specific terminology according to documentation.}
\label{tab:vendor_specific_terms}
\begin{tabular}{l l l l}
\toprule
& \multicolumn{3}{c}{\textbf{Vendor}} \\
\textbf{Term} & NVIDIA & Qualcomm Adreno & Apple AGX \\
\midrule
Smallest Execution Unit & SP/DP/SFU                 & ALU                 & ALU                                \\
Thread                  & Thread                    & Fiber               & Thread                             \\
Thread Model            & Cluster/Block/Thread group & Workgroup/Wave      & Grid/Thread group                   \\
Core Execution Unit     & 32 Threads per Warp       & 32 Threads per Wave & 32 Threads per SIMD/Warp/Wavefront \\
\bottomrule
\end{tabular}
\end{center}
\end{table*}
Since GPUs contains much more cores than the typical main processor of the system, they are well suited for highly parallelizable workloads. With \textit{General Purpose Computation for GPUs} (GPGPU), the usage of GPUs beyond their initial work scope has been proposed. First commercially introduced in 2006 by NVIDIA with their proprietary CUDA framework and later, in 2009, with the publication of first widely supported open-source GPGPU specification OpenCL, the GPGPU concept has been adopted by every major vendor. Frameworks, which have primarily been focused on graphics, such as DirectX and OpenGL, have also added support for their variants of compute shaders. Thus, every GPU in high-performance systems as well as mobile systems now supports GPGPU programming. \\
The compute GPU pipeline is much simpler than its graphics counterpart, since data passed to the GPU is normally returned to the calling application directly after shader execution finished instead of being passed through a complex multi-stage pipeline. \\
This new programming model opened the door for many opportunities: For one, GPUs have been extensively been used for physics simulations and other scientific applications relying on parallelized work. However, another specific application use case which has emerged are neural networks. GPUs allow to parallelize big matrix operations which are fundamental for neural networks. Open-source frameworks such as PyTorch and Tensorflow \cite{imambi2021pytorch} are the predominant cornerstone of modern AI, but also vendors themselves tend to provide their own optimized libraries \cite{qualcommneuralsdk}. All major frameworks have strong support for GPUs, making those the predominant accelerators for neural network workloads.

\subsection{Computing Workloads}
GPGPU APIs give the developer the most control over the underlying hardware in contrast to graphical programming interfaces. Computing programs can dispatch workloads to the GPU in the form of workitems, where each workitem is mapped to a GPU thread. All workloads are organized in a hierarchy for better scheduling and higher throughput. These hierarchies are closely related to the hardware architecture, needing the programmer to consider the underlying architecture during development. For one, leads to a higher effort needed when porting GPU programs to a newer architecture, however, this also allows for full control and maximum optimization of GPU programs on the side of the developer. \\
For computational tasks, all GPGPU APIs expose the hardware in a similar fashion. Every API provides access to multiple \textit{thread groups}, where each thread group typically consists of up to 1024 threads. One thread group is scheduled to a single core, where thread groups are then further divided in \textit{waves}, which are a set of threads that are executed concurrently in hardware on a single SIMD unit. GPU cores consist of one or more of said SIMD execution units, which can execute one sub-thread group at once. Programming interfaces allow to index a thread in a thread group via a 3-dimensional index and to further arrange thread groups in a 3-dimensional grid. These organizational scheme is present in all aforementioned interfaces. 

\subsection{Vendor Differences and Terminology}
Despite sharing a lot of similarities, custom vendor implementations still have a lot of differences, which we will highlight in this section. Especially terminology is different between vendors, with specific terms shown in Table \ref{tab:vendor_specific_terms}. NVIDIA provides the most complete documentation about their hardware, where Apple and Qualcomm document only essential information. Cores in NVIDIA GPUs contain multiple different specialized ALUs and cores for different worksets, named Raytracing Cores (RTs), Single Precision Units (SPs), Double Precision Units (DPs) and Special Function Units (SFUs). This is a contrast to the \textit{unified shader architecture}, which is utilized in Apple and Adreno GPUs, where the same execution units are responsible for all aforementioned tasks.\\
Because different vendor terminologies, we need to use generalized terminology:
\begin{itemize}
    \item \textbf{Thread}: Smallest execution unit, which corresponds to a single workitem.
    \item \textbf{Wave}: A subgroup of threads which are executed in parallel on a SIMD unit.
    \item \textbf{Thread group}: A collection of threads which is assigned to a single core.
    \item \textbf{SIMD Unit}: Execution unit in a GPU core which executes 32 threads in parallel.
    \item \textbf{GPU Core}: Contains multiple SIMD Units, a local register file and a local scheduler.
\end{itemize}
Both Qualcomm and NVIDIA specify in their documentation that they allow the execution of half-waves, where the SIMD execution unit is split up into two parts. \\ %Adreno GPUs do this on heuristical approaches, NVIDIA primarily used this form of execution up to their Fermi architecture. Half-waves are still used today in NVIDIA architectures, where the memory controller coalesces memory accesses up to half-wave size. \\
Most architectures only support a single program counter (PC) per SIMD unit and control execution via a execution mask. Thus, all threads in a wave execute the same instruction at the same time, only with different data. NVIDIA incorporated to ability to accommodate multiple PCs in a single SIMD unit, thus reducing the latency of diverging threads.
\section{Threat Model}
\label{sec:threatmodel}
Our attacks require the attacker to execute a malicious user application. Beyond this, no special privileges for the attacker application are needed.
The attacker needs to be able to execute a custom shader binary on the GPU. Modern systems provide a kernel command interface, for which an abstraction layer, such as a graphics or GPGPU library exist. The communication with the interface can either be replicated or hooked by the attacker. Modern operating systems do not consider graphic and rendering library access to be a privileged or sensitive, allowing an unprivileged attacker to execute arbitrary shader binaries.\\
In our work, we consider that both the victim and the adversary can execute arbitrary code on the same machine. On a customer device, we consider a victim application and a malicious application, where latter either aims to leak pixel or neural network data of the victim application. For cloud applications, we consider a shared GPU between multiple cloud tenants, over which the adversary aims to leak information. \\
Another assumption we make is that the adversary has knowledge about the targeted GPU application. For neural network-based attacks, we assume that the adversary at least has knowledge about what the targeted model is and has access to the model itself. This is necessary in order to correctly reconstruct the leaked data. We later see in section \ref{sec:evaluation}, that this assumption can relaxed depending on the model, where sometimes only partial information about network parameters is necessary.
\section{Uninitialized Register Access Attacks}
\label{sec:attack}

\begin{figure*}
    \centering
    \includegraphics[width=0.8\textwidth]{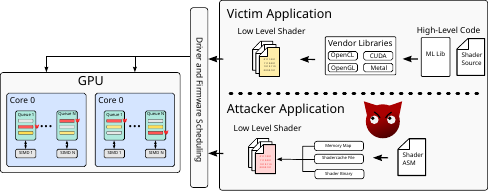}
    \caption{Overview of the attack procedure. The adversary provides hand-crafted shader kernels to the GPU interface, while normal user applications use vendor-provided interfaces. In order to achieve co-location with the targeted shaders, the attacker needs to dispatch his kernel in the execution queue of every core. }
    \label{fig:enter-label}
\end{figure*}

In this section, we present uninitialized register accesses as an information leakage mechanism targeting GPUs, allowing to leak data processed by programs executed on the GPU. First, we give a detailed insight into how the presented vulnerability works. Next, we give an overview of how to determine what data exactly is leaked by a previously executed shader, based on the knowledge of a specific shader. Lastly, we discuss several challenges that might arise when an adversary tries to leak specific data and show potential solutions on how to handle various cases.\\
\begin{figure*}
	\centering
	\begin{subfigure}{0.3\textwidth}
		\centering
		\begin{lstlisting}[language=ASM]
// Get thread number
      mov.f32f32 r0.y, r51.w
(rpt2)nop
(nop3)shl.b r0.y, r0.y, 3
(nop3)add.s r0.x, r0.x, r0.y
      mov.f32f32 r0.y, r0.x
(rpt5)nop
// Store stale register r2.x
      stib.b.untyped.1d.u32.1.imm
           r2.x
      mov.f32f32 r0.z, c8.x
		\end{lstlisting}
		\caption{Adreno kernel.} 
		\label{fig:attack_adreno}
	\end{subfigure}
	\hfill
	\begin{subfigure}{0.3\textwidth}  
		\centering 
		\begin{lstlisting}[language=ASM]
// Get thread number
get_sr r1, 
       sr80(thread_pos_in_grid.x)
// Store stale register r0
device_store 0, i32, single,
             r0, u2_u3, r1,
             unsigned, 0
stop
		\end{lstlisting}
		\caption{AGX kernel.} 
		\label{fig:attack_agx}
	\end{subfigure}
	\hfill
	\begin{subfigure}{0.3\textwidth}   
		\centering 
		\begin{lstlisting}[language=ASM]
// Get thread number
MOV  R1, c[0][28h]
S2UR UR6, SR_CTAID.X
UMOV UR7, 4
ULDC.64    UR4, c[0][160h]
UIMAD.WIDE UR4, UR6, UR7, UR4
// Store stale register R8
STG.E.SYS  [UR4], R8
EXIT
		\end{lstlisting}
		\caption{NVIDIA kernel.} 
		\label{fig:attack_nvidia}
	\end{subfigure}
	\caption{Attacker kernels for exploiting stale register reads.} 
	\label{fig:attacker_kernels}
\end{figure*}
Our attack consists of 3 steps:\\
\textbf{Step 1.} Ensure scheduling of the leakage shader after the targeted shader. \\
\textbf{Step 2.} Run the leakage shader on the GPU and copy stale register values to memory. \\
\textbf{Step 3.} Copy form GPU to host memory and reconstruct the leaked data based on its structure. \\\\
In the following, we describe how the found vulnerability behaves on different platforms, since it affects how and what kind of information is leaked. The hand-crafted kernels for general exploitation of the vulnerability are shown in Figure \ref{fig:attacker_kernels} for all three platforms. \\\\
\textbf{Adreno and AGX.} Both the underlying vulnerability as well as the exploitation process are nearly identical on Adreno and AGX. On those GPUs, registers are not cleared after shader execution, leaving possible sensitive data behind. On Adreno GPUs, up to 64 registers are available to a kernel, and each register consists of four subregisters, which can be accessed via a postfix, resulting in a total number of 256 32-bit wide registers. On AGX, the shader can claim up to 128 32-bit wide registers for computation. However, both platforms are quite unlikely to utilize all available registers in practice. \\\\
\textbf{NVIDIA.} On NVIDIA GPUs, stale register values can only be leaked if said register value has previously been written to memory via a memory store instruction. On these architectures, the memory coalescing operation is implemented on half-wave level, where the memory store is first executed for the half of a wave and for the second wave afterwards, resulting in two consecutive memory bus accesses. Since our experiments suggest that we can only leak the last written memory values, we can only leak outputs of the second half wave memory access.

\subsection{Building Blocks and Challenges}
Even though the presented vulnerability forms a powerful side-channel attack, there still arise multiple challenges which have to be solved. In the following, we will provide an overview over building blocks which are incorporated in order to form a reliable exploitable side-channel.
 Furthermore, since we leak stale data by \textit{every} running shader, one has to be able to determine which data to consider as noise and to identify only data of interest. \\\\
\textbf{Attacker Shader Creation.} The vulnerabilities we present are not necessarily exploitable from high-level code, such as Metal Compute or OpenCL. Hence, the attacker needs to craft a malicious shader binary manually. NVIDIA provides the \texttt{nvcc} compiler, which is able to compile host code and GPU CUDA C++ code into one single binary. For the compiled CUDA code, \texttt{nvcc} is able to either emit or compile parallel Thread Execution IR (PTX), which is compiled later during runtime, or to emit and compile raw GPU shader code. We use the later feature in order to maliciously modify the shader. \\
Adreno and Apple both use a different approach: Shader source code is compiled during program runtime, where the host application provides the high-level code to a vendor-provided shadered library, which in turn compiles the shader during application runtime. Therefore, the attacker first has to compile a placeholder shader and patches the binary code in memory at runtime. On Adreno GPUs, the shader binary code is held in a memory map of the \texttt{/dev/kgsl} device. The Adreno kernel driver exposes multiple memory mappings via this device, for either data, commands and code binaries passed to the GPU. This memory map are a shared memory regions with the GPU, where binary modifications are seen immediately and no further copy operations are necessary. By iterating over the memory maps, the one containing the shader binary can be found without difficulty by checking for valid disassembly. We decided to rely on OpenGL ES Compute because it is implemented on all devices with OpenGL ES 3.1 support and does not require additional libraries like OpenCL, where support is not guaranteed for every device.\\
For AGX, we employ a simliar approach: Both iOS and macOS use a shader cache file for each executed application. The shader cache is contained in a file named \texttt{functions.data} and is memory mapped when the Metal framework is loaded. Even though AGX code is placed on a shared memory region, like Adreno, we choose to modify the shader cache file since it allows for better interoperability of the exploitation technique between both OS versions. We utilize the Metal framework for shader management and modify a placeholder Metal shader. \\\\
\textbf{Scheduling.} Our goal is to leak stale register values which have been left behind by a victim shader. To achieve this, the attacker shader has to be scheduled on the same core as the victim immediately after it finished execution. However, the scheduling algorithms running in the firmware can be considered to be a blackbox to the programmer, since there is no programming interface to assign a shader to a specific GPU core. The placement of both shaders determines what and how much data can effectively be leaked. In order to understand how to effectively leak data, we need to gain a deeper insight into how the scheduling algorithms work for all platforms.\\
 Documentation for vendors state that a thread group is assigned to a single core, and that a thread group cannot be assigned to multiple cores at once. However, a core can execute multiple thread groups, even at once. Specifically for the Fermi architecture generation of NVIDIA GPUs, researchers found that shaders are assigned to cores in a round-robin fashioned manner \cite{naghibijouybari2017constructing}. We confirmed this to be the case through experiments for all GPU architectures mentioned in this paper. Consequently, the highest probability to achieve scheduling on the same core is by utilizing all cores and SIMD units available, which ensures queuing of the attacker code after the victim executed with high confidence. \\\\
\textbf{Register Remapping.} GPUs manage a large register file, where registers are allocated to running shaders by the firmware. Resultingly, the register mapping can be considered as a black-box to the adversary. While investigating register mapping on different GPUs, we found that AGX has the most agressive remapping scheme, with no observable pattern in how registers are allocated for a new shader. When two registers appear to be consecutive in one shader (e.g. \texttt{r0} and \texttt{r1}), those two registers are unlikely to appear as consecutive registers again in another shader. In constrast, Adreno has a far more predictable remapping scheme. For example, if an adversary wishes to leak the value of register \texttt{r44.y} in the victim shader, he himself needs to read from register \texttt{r44.y}, and the target value does not seem to appear in other registers. Thus, the attacker can selectively leak data of victim shaders, leading to very precise attacks on Adreno GPUs. On NVIDIA GPUs, we observed that reading from any uninitialized registers reveals nearly all data written to GPU memory. \\\\
 \textbf{Data order.} Since there are more workitems in the work queue than GPU cores available, the question arises on how the structure of data is preserved during leakage. Work is scheduled in thread groups, where all threads in a single group execute on the same core. Commonly, thread groups with a size of up to 1024 threads are typically encountered. All of these threads are split in smaller groups of the size of a single SIMD unit, which is 32 in many products, and get dispatched in parallel on a SIMD unit. We observe that we can leak data of a SIMD unit in order and further make the observation, that the waves are dispatched in order. However, since we have no knowledge of how victim thread groups are mapped to cores by the scheduler, we cannot determine the data order across thread groups. \\\\
\textbf{Determining what is leaked.} The adversary first has to determine what data is leaked from the targeted victim shaders. On NVIDIA GPUs, since only the last written values of a register during execution are leaked, intermediate values cannot be leaked. On contrary to this, AGX and Adreno GPUs leak all their register values, which conveniently can also be used to leak intermediate values of GPU calculations. \\
Moreover, the Adreno shader compiler uses high registers, such as \texttt{r51.w} often for passing runtime information to the shader In practice however, we encountered no shader which uses such high registers for any calculations. \\
Even though only register contents which are written to memory can be leaked, the intermediate layer outputs of neural network frameworks are still prone to data leakage. We examined two popular frameworks (PyTorch \cite{imambi2021pytorch} and tinygrad \cite{tinygrad}) and noticed that every layer is executed as a separate kernel. Hence, data between different layers can only be passed via memory, and all layers need to write their outputs to shared memory instead of keeping values in registers. This allows an adversary to leak intermediate layer data without much effort despite certain shortcomings of the exploitation primitive. \\\\
\textbf{Distinguishing Data from Noise.} Given the fact that GPUs process and handle extensive amounts of data per second, locating the targeted data to leak within system noise poses a unique and challenging problem. The adversary can take advantage of the structure of the data processed, depending if pixel data or different types of neural networks are targeted. Especially different neural networks are prone to leakage at different locations due to their internals. We will further investigate this challenge as well as solutions in our evaluation in section \ref{sec:evaluation}. \\
%%%%%%%%%%%%%%%%%%%%%%%%%%%%%%%%%%%%%%%%%%%%%%%%%%%%%%%%%%%%%%%%%%%%%%%%%%%%%%%%
\section{Evaluation}
\label{sec:evaluation}
In this section, we evaluate the performance of the uninitialized register reading attack under different conditions. We first present a covert channel implementation using the vulnerability, and leverage it for performance evaluation. Afterwards, we thoroughly evaluate the real-world impact of the vulnerability and how different workloads can be attacked, ranging graphics to neural network information leakage. All experiments in this section were conducted on the hardware as listed in Table \ref{tab:testedgpus}.

\begin{table}[]
	\begin{center}
		\begin{tabular}{c c c c}
			\toprule
			Work                   & Architecture        & Bandwidth & Method           \\
			\midrule
			Dutta et al. \cite{dutta2021leaky}           & Intel iGPU          & \makecell{120kb/s,\\400kb}         & \makecell{Last-Level\\ Cache}         \\ \addlinespace
			\makecell{Naghibijouybari\\et al. \cite{naghibijouybari2017constructing}} & NVIDIA              & 4Mb/s                    & \makecell{Contention\\ (ALU, memory)} \\ \addlinespace
			Dutta et al. \cite{dutta2023spy}          & NVIDIA              & 4Mb/s                    & Interconnect         \\ \addlinespace
			\makecell{Naghibijouybari\\et al. \cite{naghibijouybari2016covert}} & NVIDIA              & 400kb/s                  & \makecell{L1, L2\\ contention}        \\ \addlinespace
			Ahn et al. \cite{ahn2021network}            & NVIDIA              & 24MB/s                   & \makecell{Interconnect\\ contention}  \\ \addlinespace
			\makecell{Almusaddar\\et al. \cite{almusaddar2023exploiting}}     & Intel iGPU          & \makecell{1.65Kb/s\\4.41Kb/s}       & \makecell{Parallel\\ memory writes}   \\ \addlinespace
			Nayak et al. \cite{nayak2021mis}          & NVIDIA              & 81Kb/s                   & TLB                      \\ \addlinespace
			This work                   & \makecell{NVIDIA\\AGX\\Adreno} & \makecell{240.7Mb/s\\413.3Mb/s\\19.2Mb/s}                     & \makecell{Uninitialized\\registers}        \\
			\bottomrule
		\end{tabular}
	\end{center}
	\caption{Performance of different GPU-based Covert Channel Implementations.}
	\label{tab:ccperformance}
\end{table}

\begin{table}[!h]
	\begin{center}
		\begin{tabular}{l  c  c  c}
			\toprule
			Device             & GPU            & OS version  & Display \\
			\midrule
			Macbook Air           & M1 AGX         & \makecell{macOS 13.4.1\\Asahi Linux} & 4k \\ \addlinespace
			iPhone X           & A11 AGX        & iOS 13.6.1   & \makecell{1125$\times$\\2436} \\ \addlinespace
			Samsung Galaxy A52 & Adreno 618     & Android 13   & \makecell{1080$\times$\\2400} \\ \addlinespace
			Custom             & RTX 2070       & Ubuntu 22.04 & 4k \\
			\bottomrule
		\end{tabular}
	\end{center}
	\caption{Tested GPUs.\label{tab:testedgpus}}
\end{table}

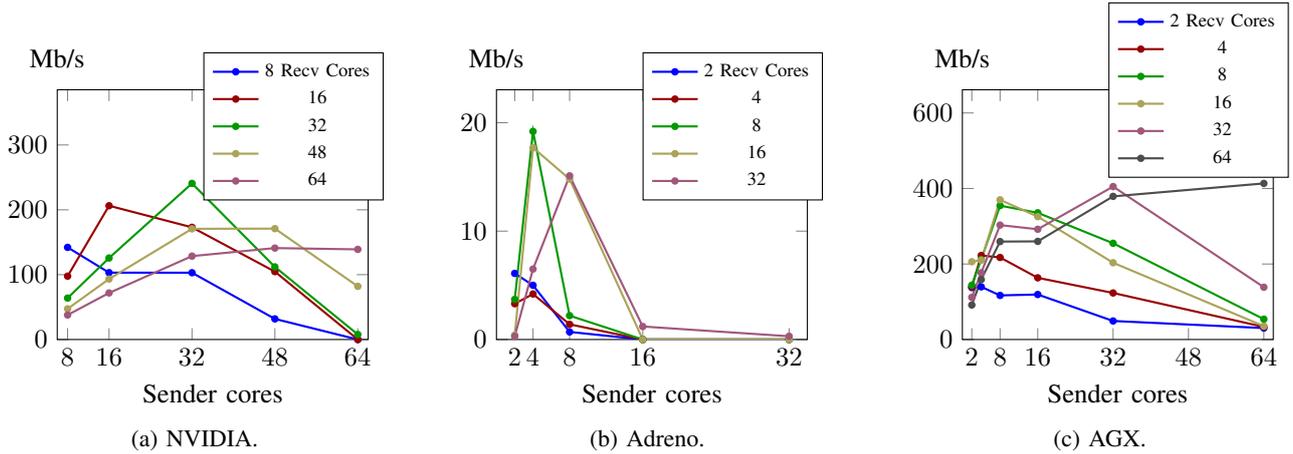
\begin{figure*}[h!]
	\centering
	\begin{subfigure}[b]{0.3\textwidth}
		\begin{tikzpicture}
			
			\begin{axis} [
				width=1.1\textwidth,
				xtick={8, 16, 32, 48, 64},
				xlabel={Sender cores},
				ylabel={Mb/s},
				every axis y label/.style={
					at={(ticklabel* cs:1.05)},
					anchor=south,
				},
				ymin=0,
				enlarge x limits = {abs = 2.0},
				enlarge y limits = {upper, value = 0.6},
				legend style= { 
					at={(1.05,1.15)},
					font = \scriptsize
				},
				every axis plot/.append style={thick}
				]
				\addplot[color=blue,mark=*,mark size=1pt] coordinates {
					(8, 142.1)
					(16, 103.1)
					(32, 103)
					(48, 31.8)
					(64, 0)
				};
				
				%\node (mark) [draw, black, circle, minimum size = 9pt, inner sep=2pt, thick] 
				%at (axis cs: 8, 142.1) {};
				
				\addplot[color=red!60!black,mark=*,mark size=1pt] coordinates {
					(8, 97.7)
					(16, 206.3)
					(32, 173)
					(48, 104.7)
					(64, 0)
				};
				
				\addplot[color=green!60!black,mark=*,mark size=1pt] coordinates {
					(8, 63.9)
					(16, 125.7)
					(32, 240.7)
					(48, 112)
					(64, 7.8)
				};
				
				\addplot[color=yellow!60!black,mark=*,mark size=1pt] coordinates {
					(8, 47.3)
					(16, 93.3)
					(32, 170.7)
					(48, 171)
					(64, 82)
				};
				
				\addplot[color=magenta!60!black,mark=*,mark size=1pt] coordinates {
					(8, 37.8)
					(16, 71.8)
					(32, 128.6)
					(48, 141)
					(64, 139)
				};
				
				\legend{8 Recv Cores, 16, 32, 48, 64}
			\end{axis}
			
		\end{tikzpicture}
		\caption{NVIDIA.}
		\label{fig:ccperformancegraph_nvidia}
	\end{subfigure}
	\hfill
	\begin{subfigure}[b]{0.3\textwidth}  
		\begin{tikzpicture}
			
			\begin{axis} [
				width=1.1\textwidth,
				xtick={2, 4, 8, 16, 32},
				xlabel={Sender cores},
				ylabel={Mb/s},
				every axis y label/.style={
					at={(ticklabel* cs:1.05)},
					anchor=south,
				},
				ymin=0,
				enlarge x limits = {abs = 2.0},
				enlarge y limits = {upper, value = 0.2},
				legend style= { 
					at={(1.05,1.15)},
					font = \scriptsize
				},
				every axis plot/.append style={thick}
				]
				
				\addplot[color=blue,mark=*,mark size=1pt] coordinates {
					(2, 6.1)
					(4, 5)
					(8, 0.7)
					(16, 0)
					(32, 0)
				};
				
				\addplot[color=red!60!black,mark=*,mark size=1pt] coordinates {
					(2, 3.3)
					(4, 4.2)
					(8, 1.4)
					(16, 0)
					(32, 0)
				};
				
				\addplot[color=green!60!black,mark=*,mark size=1pt] coordinates {
					(2, 3.7)
					(4, 19.2)
					(8, 2.2)
					(16, 0)
					(32, 0)
				};
				
				\addplot[color=yellow!60!black,mark=*,mark size=1pt] coordinates {
					(2, 0.4)
					(4, 17.7)
					(8, 14.8)
					(16, 0.0)
					(32, 0)
				};
				
				\addplot[color=magenta!60!black,mark=*,mark size=1pt] coordinates {
					(2, 0.3)
					(4, 6.5)
					(8, 15.1)
					(16, 1.2)
					(32, 0.3)
				};
				
				\legend{2 Recv Cores, 4, 8, 16, 32}
			\end{axis}
			
		\end{tikzpicture}
		\caption{Adreno.}
		\label{fig:ccperformancegraph_agx}
	\end{subfigure}
	%\vskip\baselineskip
	\hfill
	\begin{subfigure}[b]{0.3\textwidth}   
		\begin{tikzpicture}
			
			\begin{axis} [
				width=1.1\textwidth,
				xtick={2, 8, 16, 32, 48, 64},
				xlabel={Sender cores},
				ylabel={Mb/s},
				every axis y label/.style={
					at={(ticklabel* cs:1.05)},
					anchor=south,
				},
				ymin=0,
				enlarge x limits = {abs = 2.0},
				enlarge y limits = {upper, value = 0.6},
				legend style= { 
					at={(1.05,1.35)},
					font = \scriptsize
				},
				every axis plot/.append style={thick}
				]
				\addplot[color=blue,mark=*,mark size=1pt] coordinates {
					(2, 137.5)
					(4, 139.5)
					(8, 116.8)
					(16, 119.2)
					(32, 49.1)
					(64, 30.3)
				};
				
				\addplot[color=red!60!black,mark=*,mark size=1pt] coordinates {
					(2, 139.5)
					(4, 222.6)
					(8, 217.3)
					(16, 163.4)
					(32, 123.3)
					(64, 33.3)
				};
				
				\addplot[color=green!60!black,mark=*,mark size=1pt] coordinates {
					(2, 143.8)
					(4, 212.1)
					(8, 354.6)
					(16, 335.6)
					(32, 254.9)
					(64, 53.9)
				};
				
				\addplot[color=yellow!60!black,mark=*,mark size=1pt] coordinates {
					(2, 206)
					(4, 210.1)
					(8, 370)
					(16, 325.5)
					(32, 203.3)
					(64, 35.4)
				};
				
				\addplot[color=magenta!60!black,mark=*,mark size=1pt] coordinates {
					(2, 111.7)
					(4, 176.7)
					(8, 303)
					(16, 292)
					(32, 405.1)
					(64, 138.5)
				};
				
				\addplot[color=gray!60!black,mark=*,mark size=1pt] coordinates {
					(2, 91.4)
					(4, 158.9)
					(8, 259.4)
					(16, 259.9)
					(32, 379.1)
					(64, 413.4)
				};
				
				\legend{2 Recv Cores, 4, 8, 16, 32, 64}
			\end{axis}
			
		\end{tikzpicture}
		\caption{AGX.}
		\label{fig:ccperformancegraph_adreno}
	\end{subfigure}
	\vspace*{.3cm}
	\caption{Covert Channel Performance evaluation for varying numbers of thread groups for sender and receiver.} 
	\label{fig:ccperformancegraphs}
\end{figure*}

\subsection{Covert Channel}
We first introduce how the uninitialized register read exploit can be utilized for creating a covert channel. On Android and iOS, where only a given API can be used for Inter-Process Communication (IPC), attackers can aim to use covert channels in order to evade either restrictions of IPC usage or evade being detected. Since GPU access is unprivileged on both operating systems, stale GPU register values are an ideal building block for covert channels. \\
As we can read stale data, the bandwidth and the noise resistance of the covert channel is high. On Adreno GPUs, mostly registers \texttt{r0} and \texttt{r1} are used, even though up to 64 general-purpose registers are available. Thus, the usage of higher registers for a covert channel is advised, as they are less likely to be used and overwritten in our observations. \\
To covertly transmit data, the sender executes a simple shader, which writes the passed data to send to a register of choice. To uniquely identify the sender data, a format as \texttt{ magic\_value | counter | data } is used. If the magic value and counter are e.g. 4 bytes in size, 124 bytes (on AGX and Adreno) or 60 bytes (on NVIDIA) can be transmitted respectively. Based on the magic value and the counter, the receiver of the covert channel can easily recover the original message. Performance evaluation results for the covert channel implementation compared to raw data leakage speeds are shown in Table \ref{tab:performancerawvscc}. Our covert channel can achieve up to  413.3Mb/s on AGX GPUs, 240.7Mb/s on NVIDIA, and 19.2Mb/s on Adreno, which is drastically more than other proposed GPU-based covert channel mechanisms. Most GPU-based covert channels unveiled by researchers are able to leak data in ranges of Kb/s to Mb/s. A detailed comparison with other research is presented in table \ref{tab:ccperformance}. To the best of our knowledge, stale register reads is the first GPU-based covert channel which achieves a bandwidth in the range of hundreds of Mb/s.

\subsection{Performance}
By building a covert channel, we developed a good base for estimating an upper bound for how much information can be leaked when targeting one specific application. Since the covert channel setup allows to measure the transfer rate between two applications without any system noise, we can measure the maximum rate at which data of one victim application can be extracted. \\
The raw leakage rate is much larger on all systems, mainly due to the GPU also rendering the display contents. In order to also evaluate the possibility of contention, we further showcase the leakage performance in detail with varying amounts of workgroups dispatched to the GPU, as depicted in Figure \ref{fig:ccperformancegraphs}. 

For both AGX and NVIDIA, the configuration where maximum throughput is achieved always happens when the number of sender cores matches the number of receiver cores. On NVIDIA, the highest throughput is reached when both the receiver and sender use 32 thread groups, whereas on AGX the maximum is achieved with 64 thread groups for both. Using more sender then receiver thread groups and vice versa leads to under-utilization of the covert channel. An interesting situation happens on NVIDIA GPUs, if 64 sender groups, but 8 or 16 receiver groups are utilized: We observed that the black-box scheduler does not schedule the sender and receiver in such a way, that any leakage can be observed. \\
\iffalse
\begin{figure}[h!]
	\begin{center}
		\begin{tikzpicture}
			
			\begin{axis} [
				width=.45\textwidth,
				height=.3\textwidth,
				xbar,
				bar width = 7pt,
				xmin = 0,
				xlabel= GB/s,
				yticklabels = {AGX\\64$\times$64, NVIDIA\\8$\times$8, Adreno\\4$\times$8},
				ytick = {0,1,2},
				yticklabel style = {rotate = 20, anchor = east, align = center},
				enlarge x limits = {abs = .8},
				enlarge y limits = {upper, value = 0.6, abs=3*\pgfplotbarwidth},
				legend style= {
					at={(1.05,1.15)},
					font = \footnotesize 
				}
				]
				\addplot[fill=blue!30] coordinates {
					(82, 0)
					(6.3, 1)
					(0.130, 2)
				};
				
				\addplot[fill=red!30] coordinates {
					(0.43, 0)
					(0.24, 1)
					(0.02, 2)
				};
				
				\legend{Raw Leakage Performance, Covert Channel Performance}
			\end{axis}
			
		\end{tikzpicture}
	\end{center}
	\caption{Covert Channel performance in comparison to raw data leakage performance.}
	\label{fig:ccperformance}
\end{figure}
\fi

\begin{table}[h!]
	\begin{center}
		\begin{tabular}{lllll}
			\toprule 
			Vendor & \makecell{Sender $\times$\\Receiver} & CC & Raw & Ratio \\
			\midrule
			NVIDIA  & \makecell{32$\times$32\\32$\times$64}    & \makecell{240Mb/s\\7.8Mb/s}   & \makecell{4.4Gb/s\\4.06Gb/2}    & \makecell{0.053\\0.0019}      \\ \addlinespace
			Adreno  & \makecell{8$\times$4\\32$\times$16}    & \makecell{19.2Mb/s\\1.2Mb/s}   & \makecell{246Mb/s\\459Mb/s}    & \makecell{0.078\\0.0026}      \\ \addlinespace
			AGX     & \makecell{64$\times$64\\2$\times$64}    & \makecell{413.4Mb/s\\30.3Mb/s}   & \makecell{82.2Gb/s\\2.81Gb/s}    & \makecell{0.0049\\0.0105}      \\ \addlinespace
			\bottomrule
		\end{tabular}
	\end{center}
	\caption{Covert Channel performance in comparison to raw data leakage performance. Best-case with worst-case configurations are compared.} 
	\label{tab:performancerawvscc}
\end{table}

Out of all vendors, we observed the slowest leakage on Adreno. We attribute this to the usage of OpenGL Compute, a very rarely used GPGPU implementation which is nonetheless widely supported. We identified the bottleneck to be the shader creation, not its execution time. As such, we accomplished covert channel speeds of only up to 19.2Mb/s. Nevertheless, our implementation on Adreno is still faster than many other covert channel implementations. We also observed on Adreno that the usage of many cores quickly leads to a bottleneck on the device. The fastest configuration uses 4 sender and 8 receiver cores.\\
%The amount of system noise, which contains data which is irrelevant for the receiver or attacker, is also of interest to us. As depictured in Figure \ref{fig:ccperformancegraphs}
This performance evaluation also provides an understanding of how many cores an adversary should utilize if he wishes to leak data with a high rate from a victim shader. On both NVIDIA and AGX, he needs to dispatch the same number of thread groups as the victim to achieve the best results.

\subsection{Leaking Fragment Shader Data}
\begin{figure}
    \centering
    \includegraphics[width=0.4\textwidth]{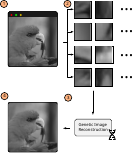}
    \caption{Fragment shader leaking approach. Leaked fragments are reconstructed via a genetic puzzle solving algorithm, which is able to reconstruct the screen data.}
    \label{fig:fragment_reconstruct}
\end{figure}

Modern operating systems employ a strict permission system, which manages app access to sensitive resources. One sensitive resource is the permission to record the screen, whereby the OS requests explicit user consent before an application can record the user's screen. In this section, we showcase how a malicious app can circumvent this protection mechanism by targeting fragment shaders. \\
The fragment shader is the one render pipeline step responsible for producing the color data of rendered primitives. If an adversary wishes to leak pixel information on the target system, he therefore has to target fragmentation shader outputs. Mobile GPUs often employ a rendering strategy called \textit{tiled rendering}. Because of limited memory space and bandwidth, the screen is partitioned into multiple tiles, and one tile is rendered by a single core. Afterwards, the results are stored in a local tile memory instead of directly being written to the framebuffer, which resides in the  bandwidth-constrained main memory of the system. As such, the tile size is primarily determined by the tile memory size. \\
\begin{lstlisting}[language=ASM, caption={Simplified fragment shader disassembly for AGX GPUs. After texture samping, registers r0-r3 contain pixel information which is stored in tile memory afterwards.}, captionpos=b, label={listing:frag_disasm}, frame=single]
// texture sampling
texture_sample   0, r1l_r1h_r2l, ts0, ss0, tex_2d
wait             0
fadd32           r0, r1l.discard, -0.0 // red
fadd32           r1, r1h.discard, -0.0 // green
fadd32           r2, r2l.discard, -0.0 // blue
mov_imm          r3, 0x3f800000        // alpha
writeout         12, 0
st_tile          r0_r1_r2_r3, quad  // Write RGBA to output
stop 
\end{lstlisting}
\textbf{Reverse Engineering Tiling.} Before pixel data can effectively be leaked from registers, we first have to gain an insight into how tiled rendering is implemented on modern GPUs. Tiles are rendered on a per-application basis, meaning that only the window contents of the rendering applications are divided into tiles instead of the whole screen. Thus, if we are able to leak tile data, the leaked tile will only contain pixel data of one application only. First, we need to understand how screen pixels are mapped to tiles. In order to understand the hardware pixel mapping process, we first develop a fragment shader which embedds pixel coordinates in the pixel color. Leaking this data, we can extract the pixel to SIMD execution unit mapping, allowing us to later correctly reconstruct leaked tiles. Moreover, we also made another observation during the reverse engineering process: Not only is pixel color data represented as a floating point between $0.0$ and $1.0$, but additionally the other parts of the rendering pipeline work on data in that range. As an example, the vertex on-screen positions are also in the same value range, which results in another challenge for pixel reconstruction and requires further attention. \\\\
\textbf{Leaking Multiple Tiles.} When leaking multiple tiles, the original input image is not trivial to reconstruct due to the vast amount of leaked small tiles. \textit{Jigsaw puzzle solving} algorithms are a subclass of computational imaging algorithms, which aim to solve such image puzzle problems for arbitrary input images \cite{pomeranz2011fully}. In our work, we employ the algorithm proposed by Sholomon et al \cite{sholomon2016automatic} due to its ability to recover image puzzles with up to 30.000 pieces. The presented algorithm is based on a Genetic Algorithm (GA) approach, which crosses over two parent inputs. Each of the parent inputs is an unfinished arrangement of the tiles, which get processed by a crossover operator. The presented operator is able to convey arrangement information from the parents to the child, if subparts of the image are already solved, thus only keeping suitable information of the parents and rejecting false tile arrangements. Figure \ref{fig:fragment_reconstruct} depictures the adversary's procedure for leaking data.
First, the victim application has to render its contents on screen in order to leave stale fragment data in registers. Afterwards, the adversary has to read out all of this fragment data. This step requires to properly separate system noise from target data. As shown in the example fragment shader disassembly in Listing \ref{listing:frag_disasm}, register \texttt{r3} sets the alpha value of the rendered texture to $1.0$. Fragments can thus be properly identified if there are uniform distributed leaked values of $1.0$ in the leaked data, which further occur around three times as much as other seen values, since registers \texttt{r0-r2} contain color information. Empirical evaluation shows that this is a reliable approach to identify rendered fragments. Due to register remapping, other leaked values cannot be properly assigned to a single color. We thus choose to reconstruct the data in grayscale color mode, by eliminating fragments with a high similarity, resulting in a single channel of color. \\\\

\subsection{Leaking Data from Convolutional Neural Networks}
\begin{figure}
    \centering
    \begin{subfigure}[t]{0.2\textwidth}
        \centering
        \includegraphics[width=\textwidth]{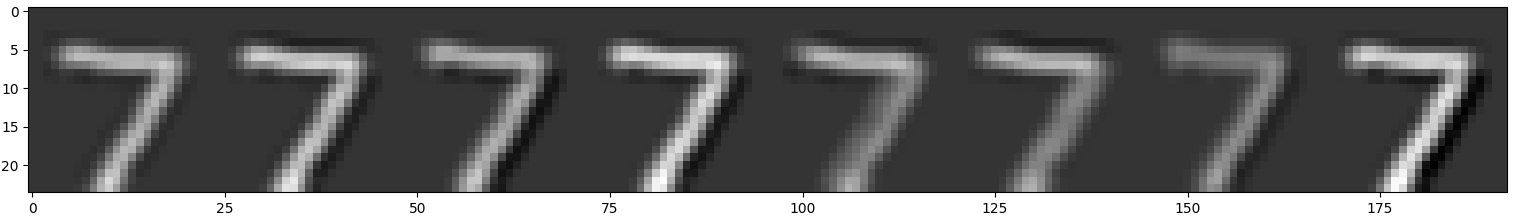}
        \caption{Original convolutional layer output.}
        \label{fig:cnn_example_output}
    \end{subfigure}
    \hfill
    \begin{subfigure}[t]{0.2\textwidth}   
    	\centering 
    	\includegraphics[width=\textwidth]{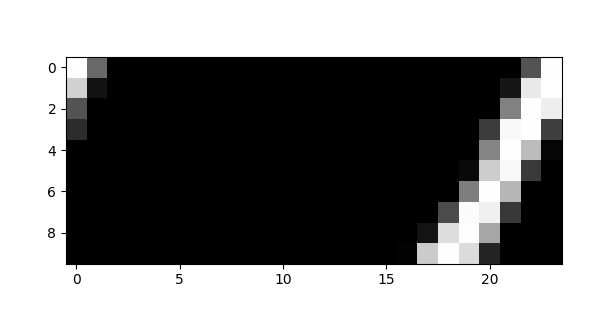}
    	\caption{\small{Consecutively leaked data on Adreno.}} 
    	\label{fig:cnn_leak_adreno}
    \end{subfigure}
    \vskip\baselineskip
    \begin{subfigure}[t]{0.2\textwidth}  
    	\centering 
    	\includegraphics[width=0.5\textwidth]{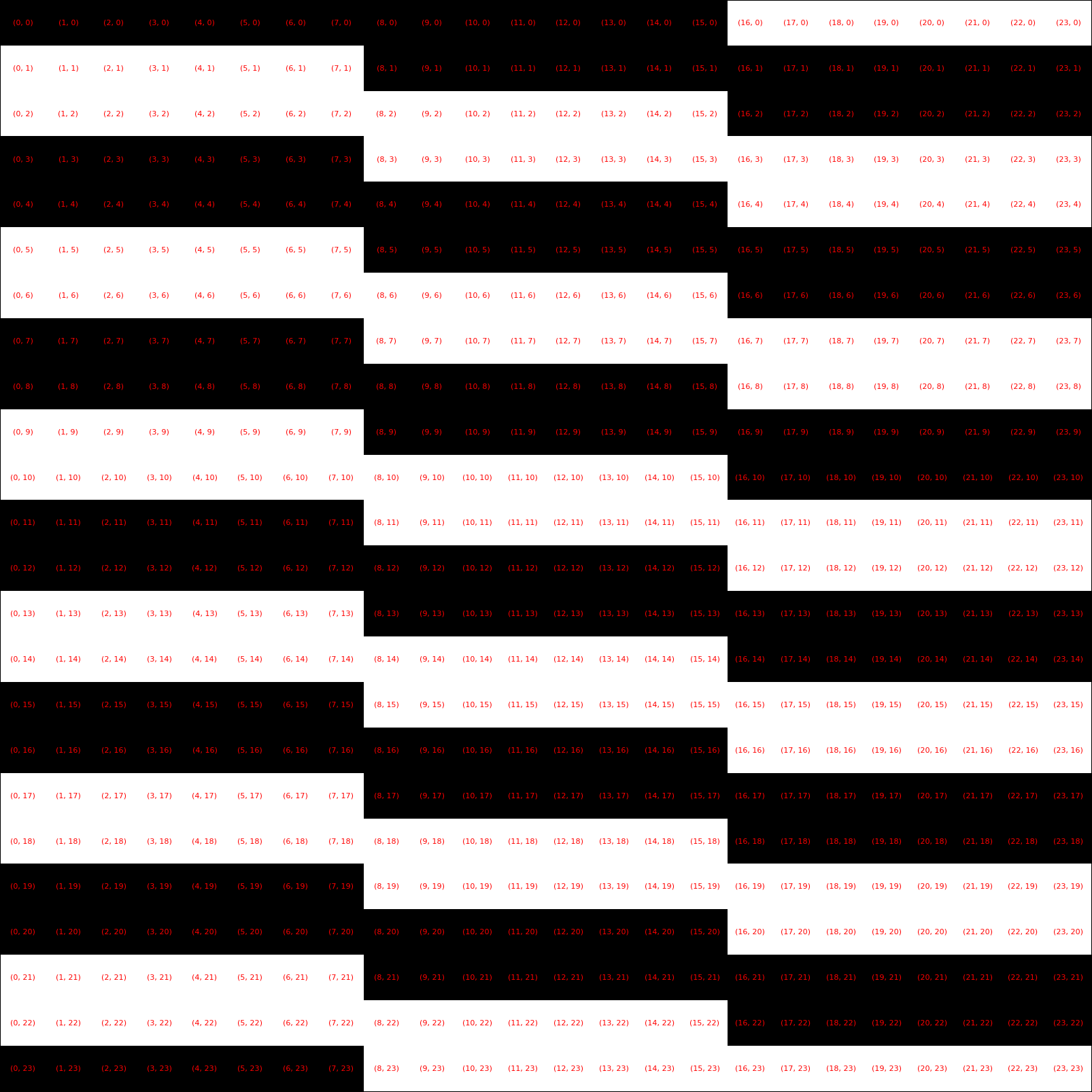}
    	\caption{\small{Leak pattern on NVIDIA. White pixels indicate that these CNN pixel outputs can be successfully leaked.}}
    	\label{fig:cnn_leak_pattern}
    \end{subfigure}
    \hfill
    \begin{subfigure}[t]{0.2\textwidth}   
    	\centering 
    	\includegraphics[width=0.5\textwidth]{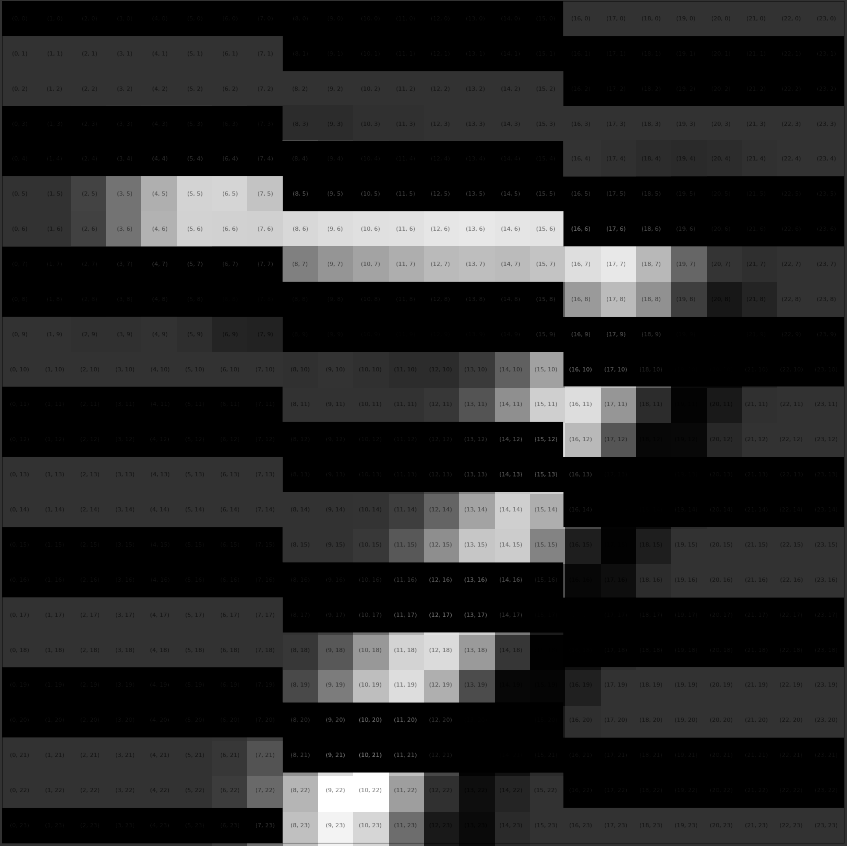}
    	\caption{\small{Heuristically reconstructed data on NVIDIA.}}
    	\label{fig:cnn_reconstructed_nvidia}
    \end{subfigure}
    \vspace*{.3cm}
    \caption{CNNs data leakage.} 
    \label{fig:cnn_leak}
\end{figure}

Convolutional Neural Networks (CNNs) are networks, whose architecture and connectivity is inspired by the animal visual cortex. By using convolutions, specialized layers apply filters to the input, in order to automatically learn features of images. Albeit being primarily focused on visual tasks, CNNs can also be used for processing other structured data, and have been adapted to text and audio. \\
CNNs find application in a variety of image-related tasks, such as object detection, classification and feature extraction. Typical building blocks for CNNs are convolutional, pooling and fully connected layers. Convolutional layers act as automatically learned image filters, extract high-level features of an input image. Pooling layers down-sample the image as it passes through the network. Different pooling-functions, such as the maximum or average pooling, multiple pixels of the layer input are mapped to a single pixel on the output, thus reducing the image size. Pooling layers fulfil multiple purposes, from retaining only the most important features in an area, to reducing the computational cost. Typically, the last layer of a CNN is a fully connected layer, with as many neurons as classes to detect. \\
When attacking CNNs, we specifically target the output of the first convolutional layer. An example output of said layer is shown in Figure \ref{fig:cnn_example_output}. Since CNNs apply multiple filter kernels to the input image, the first output therefore contains the image multiple times. Henceforth, the convolution outputs can give a very strong indication of what the original input values were. Our implementation, which is based on the original authors design \cite{o2015introduction}, consists of four layers: Convolutional, RELU, max pooling and a fully connected layer. Our implementation works on input images of size 28 $\times$ 28. The targeted convolutional layer processes its inputs in 8 thread groups with 576 threads each, and each filter kernel is applied to the input image in parallel in a separate thread group. When attacking the CNN, we consider a realistic scenario, where the CNN in executed only once. \\\\
\textbf{Leaking Data on NVIDIA.} On NVIDIA GPUs, we can leak 16 consecutive pixels per single kernel invocation due to memory coalescing. Further examination revealed that we can only leak the first 16 outputs of a warp of size 32, resulting in a leakage pattern as depictured in Figure \ref{fig:cnn_leak_pattern}. One attacker thread group has shown to be capable of recovering one half-wavefront, resulting in the challenge to reconstruct the leaked image from the leaked data. Thus, we implemented a heuristic approach to reconstruct the leaked output: As depicted in the leak pattern, the outputs of leaked half-wavefronts overlap by 8 pixels. Image reconstruction is thus be done by arranging the leaked data such that the pixel difference between two half-wavefronts is minimal at the overlapped region. The resulting reconstruction of the leaked image is shown in Figure \ref{fig:cnn_reconstructed_nvidia}. \\\\
\textbf{Leaking Data on Adreno.} Experiments revealed that the probability of leaking consecutive data is higher on Adreno due to a less aggressive register remapping strategy. By setting the thread group size to the same one as of the victim, we can leak 256 consecutive pixel values without any need of reordering or any image reconstruction algorithms. Further, we observed that leaked data is always led by a fixed number of zero bytes and registers are less aggressively remapped, resulting in straightforward identification of the leaked image.
%\begin{figure}
%    \centering
%    \includegraphics[width=0.18\textwidth]{images/adreno_cnn_leak.png}
%    \caption{Leaked convolutional layer output on Adreno.}
%    \label{fig:adreno-cnn-leak}
%\end{figure}
A single result of a leaked value is depictured in Figure \ref{fig:cnn_leak_adreno}. Our attacker targets an intermediate filter result, which is held in register \texttt{r2.x}, which results in the slightly different image. Parts of the original first layer output are clearly visible. Since CNNs mostly process images, even a distorted image contains relevant information. In a realistic scenario where the victim CNN has only been executed once, an attacker is able to capture up to 44.4\% of one convolution first layer output. Another insightful implication of this is that data on Adreno can be leaked across multiple thread groups in order.

\subsection{Attacking Large Language Models}
Recently, Large Language Models (LLMs) rapidly gained a lot of popularity. LLMs are capable of a variety of language-based tasks, ranging from more simple tasks like sentiment analysis, summarization and translation to complex chain-of-thought tasks \cite{wei2022emergent}. Their versatility leads to their usage for a variety of use-cases, and companies offer convenient programming interfaces for developers to integrate the power of LLMs into their own services for low costs. Besides handling user support tasks such automatically handling customer requests, LLMs are being used as a supportive tool at the workplace. LLM vendors even begin to offer enterprise editions of their products, which promise to not use enterprise customer's data for LLM learning purposes. As such, LLMs handle a huge variety of potentially confidential and highly personal information. \\
In this section, we present how an attacker can extract information from a GPT-2 LLM running on the same GPU. In this setup, we target a NVIDIA GPU, since the model parameter size explosion of LLMs can only be handled by thoses datacenter-class accelerators. 

\subsubsection{GPT-2 Architecture}
\begin{figure}
    \centering
    \includegraphics[width=0.28\textwidth]{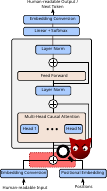}
    \caption{Simplified GPT-2 architecture. The attacker targets the output of the positional embedding of the input token embeddings.}
    \label{fig:transformer}
\end{figure}

The recent rise of LLMs has been thanks to the Transformer architecture, a neural network architecture presented by Google researchers, which is depictured in Figure \ref{fig:transformer}. We specifically target the second version of the Generative Pre-trained Transformer by OpenAI, GPT-2, which is the open-source predecessor to the proprietary model used in ChatGPT \cite{radford2019language, openai2023gpt4}.  \\
Like other text-processing neural networks, the input is first mapped to an embedding space, which is a vector representation of tokens. As such, semantically similar tokens are spatially near each other in this vector space. When giving a human-readable input, a tokenizer first has to convert this input to a vector in embedding space. GPT-2 particularity uses Byte Pair Encoding (BPE). BPE works on the byte-level instead of a given character encoding, such as ASCII. \\
The key component to this architecture is the attention mechanism, which allows the network to weigh the importance of different parts of the sequence when generating words. For each word in the sequence, three vectors are provided: query (q), key (k) and value (v). When generating a word, the attention mechanism calculates similarity scores between different parts of the sequence using these vectors. A token can advertise its features via the value and search for matching tokens using the query and key parameters. As a result, the LLM can learn which words are of relevance for each word, enabling it to process and generate long sequences of data. This is a huge advantage in contrast to other, preestablished text processing networks, such as Recurrent Neural Networks (RNNs), which are shown to have difficulties processing long input sequences due to vanishing gradients during learning. The GPT-2 architecture further involves multiple attention mechanisms in parallel - this allows for learning multiple features of word dependence. Not only are semantic connections between words captured, but attention applied in parallel allows for capturing different grammar rules in different heads. \\
The last important layer in the Transformer is a Feed-Forward Neural Network, which is a 2-layer fully connected network. This adds further trainable parameters to the network, which allow each token to process the previously seen similarities.

\subsubsection{Information Extraction}
Our attack primarily focuses on extracting information from the first calculation of the GPT-2 architecture, where embeddings and positional information are encoded. Embeddings allow to encode tokens as vectors, allowing a neural network to leverage additional information encoded in these vectors. The embedding space is represented by a matrix of size 50257 $\times$ 768, containing a total number of 50257 possible tokens, where each one represented by a vector with a dimension of 768. GPT also employs positional encodings, which, contain information about the position of a token. Since the order of words is an important aspect of language, LLMs leverage this way of encoding the order of words for learning how to process words in the correct order. \\
In this work, we focus on the implementation of GPT-2 provided by the tinygrad framework \cite{tinygrad}, because it contains minimal boilerplate code and allows for rapid changes to the framework code, which allowed us to easily debug internals of the GPT-2 implementation. Our attack targets the first step of GPT-2, where tokens are positionally encoded and feed into the Transformer architecture. Since tokens are encoded using a fixed space, each embedding can only be encoded to a fixed number of positions. GPT-2 has a limited context length of 1024, resulting in $50257 \cdot 1024 = 51463168$ possible vectors, which the result of the encoding can be, thus limiting the search space. \\

The primary advantage of targeting the embedding layer is that the embedding layer is independent of the model size. This allows an adversary to target even other LLMs, with billions of parameters, since all LLMs use a variant of an embedding scheme, which allows to scale the attack even to the biggest models available. \\\\
\textbf{Efficient Reconstruction.} For reconstructing the leaked input/output of GPT-2, we implemented an efficient reconstruction algorithm, which is a necessity since a brute-force search approach exceeds the memory limits on our machine. First, we calculate all possible permutations of embeddings and positional encodings (Line \ref{line:algo1}). We use the values of all created vectors as an index for look-up-table, where each entry determines whether a value is contained in the matrix and if yes, for which tokens and positions (Line \ref{line:algo2}). We use the binary integer representation of the floating point values for the table index. We afterwards iterate over all leaked data in chunks of size 16 (Line \ref{line:algo3}). Now, for all elements in this chunk, we determine if they are part of a embedding/position permutation an record which we encountered how often (Line \ref{line:algo4}). We record the index of the currently computed value in an embedding. We do this in order to check if we found 16 consecutive values in an embedding vector (Line \ref{line:algo5}). If that is the case, we can determine that the current chunk is an embedding at a certain position. Our experiments revealed that the reconstruction accuracy is 100\%, and we never encountered falsely reconstructed or missing data.\\
\begin{lstlisting}[language=algorithm, caption={Efficient LLM output reconstruction algorithm.}, captionpos=b, mathescape=true, frame=single, escapechar=|]
data: float[]
max_pos: int

embeddings: float[50257, 726]
pos_embeddings: float[50257, 726]

begin
  lut $\gets$ []
  llm_outputs $\gets$ []
  
  // Build LUT
  for pos = 0 : max_pos: |\label{line:algo1}|
    for embedding : embeddings:
    token_at_pos $\gets$ embedding + 
                   pos_embeddings[pos]
  
    for v in token_at_pos: |\label{line:algo2}|
      // Use integer representation as index
      lut[(int)v] $\gets$ (pos, embedding)
  
  // Lookup
  for idx = 0 : length(data); idx+=16: |\label{line:algo3}|
    chunk $\gets$ data[idx:idx+16]
    for embedding in embeddings:
      ctr $\gets$ 0
  
      emb_indices $\gets$ []
      for v in chunk: |\label{line:algo4}|
        // Use integer representation as index
        emb, pos $\gets$ lut[(int)v]
        if emb, pos $\neq$ ($\O$,$\O$):
          ctr++
          // Determine at which index value v 
          // is in the embedding
          $idx_{v\_in\_emb}$ $\gets$ get_idx_in_emb(v, emb, pos)
          emb_indices $\gets$ [
              emb_indices, $idx_{v\_in\_emb}$]
  
      if ctr $\geq$ 16 $\land$  |\label{line:algo5}|
         ascending(embedding_v_indices):
        llm_outputs.append(embedding)
        break
end
\end{lstlisting}
On our machine (8-core AMD Ryzen 7 3800x, 32GB RAM), the LUT table consumes around 20GB memory and token reconstruction speed is 22.5 seconds/token. Calculating the LUT is a one-time effort, since it can conveniently be stored on disk after creation. Calculating the LUT for up to 30 token position takes 8 minutes.
\section{Countermeasures}
\label{sec:countermeasures}
In this section, we discuss possible countermeasures to thwart exploitation of the presented vulnerability. At first, we discuss a userleve-based countermeasure and two countermeasure approaches on the operating system level next. Afterwards, we discuss the firmware-based approach.\\\\
\textbf{Cleanup after Execution.} The self-protective approach, where a kernel tries to protect itself by inserting additional instructions at the end to zero used does only work on AGX and Adreno GPUs. Since shader compilation is only accessible via an abstract interface, this would require either modification of vendor-provided libraries or a runtime-based protection mechanism, which builds upon the shader modification techniques as presented in section \ref{sec:attack}. This gives application developers which want to confidentially process data on the GPU a simple way of implementing a countermeasure against the shown attacks. Therefore, this approach is a good alternative to operating system-based mitigations, which might not be available on end-user devices, since they require modification and updates of low-level components. \\\\
\textbf{Compartmentalized Compilation.}
The vulnerability presented in this paper crosses application boundaries to applications by executing a \textit{maliciously crafted shader binary} on the GPU. As such, a solution is to compartmentalize the shader source compilation in a privileged process with a tightly controlled and abstract interface. Instead of executing the compilation process inside the user application, we propose a privileged system service, a \textit{privileged shader compiler}, which handles the compilation process. This privileged compiler passes the resulting binary to the kernel and returns a handle to the calling application. Special care needs to be taken to apply proper security and authentication measures to the handle, such that applications cannot execute the compiled shaders of other applications. However, this approach suffers by bloating the Trusted Computing Base (TCB) of the underlying system, by inserting a whole compiler toolchain. \\\\
\textbf{OS Kernel-based Sanitizer.}
 Another approach is to sanitize all compiled shaders before execution on the kernel side. We propose a kernel-based sanitization module, which disassembles every shader prior to its execution. Afterwards, the sanitizer iterates over all disassembled instructions and checks if any uninitialized registers are read. If this is the case, the the shader is rejected for execution. \\
We argue that this approach is far more lightweight than the compartmented shader compilation technique. However, it cannot be built from existing codebases and needs to be developed from the ground up with the kernel environment in mind. Because most available shader compilers and disassemblers, both open-source and proprietary, use LLVM internally, porting is unfeasible. For one, the size of the LLVM framework would blow up the kernel's size tremendously and might introduce a larger attack surface for software-based attacks against the kernel just as for the compartmentalized compilation approach. For another, LLVM-based disassemblers are not designed with speed in mind, which poses a crucial bottleneck for system performance. Handwritten disassemblers are mostly small in size and therefore the best solution for a software-based approach. \\\\
\textbf{Firmware Modification.}
Lastly, this type of attack can also be mitigated on the GPU firmware side. As the most trivial solution, the GPU firmware can be modified to clear the contents of a newly allocated register window, mitigating most of the attacks presented in the paper. However, this approach requires development effort on the vendor's side. GPUs typically include secure boot mechanisms for their management firmware, which prevents loading unauthenticated firmware images. As such, this approach cannot be implemented by any other entities than the GPU vendor.
%%%%%%%%%%%%%%%%%%%%%%%%%%%%%%%%%%%%%%%%%%%%%%%%%%%%%%%%%%%%%%%%%%%%%%%%%%%%%%%%
\section{Discussion}
In this section, we discuss various aspects of our attack and evaluation and present possible future work afterwards. \\\\
\label{sec:discussion}
\textbf{Bandwidth.} Since reading stale register values in a shader is not compute-intensive, the data bandwidth for data leakage is constrained by memory speed. On Adreno however, we detected that the leakage is mostly constrained by the vendor shader dispatch code due to our usage of OpenGL Compute. Leakage bandwidth  We aim for our attacks to be as stealthy as possible in order to avoid detection by users on the system. An adversary must be aware of resource contention occurring on the system and might need to lower the interval of kernel executions. For all data leakage attacks, we first read stale registers and simply return the leaked register contents via a shared memory location to the host. We later resume any analysis on the host side, because data analysis on the shader side leads to long shader runtimes, which effectively influences GPU performance much more, making our attacks noticeable to system users. We aim for our attacks to be stealthy and not noticeable for the user. For example, Adreno documentation directly states a maximum shader runtime of 10ms, which should not be exceeded in order to not influence Android GUI rendering, which occurs every 30ms. Estimating an upper bound of data which can be leaked gives us an estimation on how practical these attacks are. A faster shader execution time also gives us the advantage that more shader executions can be dispatched, thus occupying more entries in the scheduler pipelines and thus increasing the chance that the attacker shader is executed diretly after the targeted victim shader.\\\\
\textbf{Limitations.} One limitation, as mentioned previously, is only data which is processed within a single SIMD unit can be leaked in order. Despite this limitation, we presented solutions for various workloads. However, when leaking unstructured data, the problem still remains. Another limitation is the exploitataility from High-Level Code. Our investigation shows that binary shader access is needed in all scenarios for enforcing reading stale register values. All shader compilers on the investigated platforms did not emit stale register reads when e.g. reading an uninitialized variable in high-level shader code. LLVM is the main framework used for shader compilation and supports default variable initialization for many years. We explored Javascript WebGL implementations of three major browsers (Chrome, Firefox and Safari) and found that those implementations initialize registers to zero before usage. \\\\
\textbf{Dedicated Neural Accelerators.} Apple devices contain a dedicated Neural Engine (ANE) since the introduction of the iPhone X. Frameworks provided by Apple, which abstract the underlying devices (such as the VisionFramework), often use CoreML under the hood. CoreML can automatically decide with which accelerator backend a neural network should be executed and chooses between CPU, GPU and ANE. This decision process is only partially documented, for example, Apple states that networks needing 32 bit floaing-point numbers are executed only on the CPU and GPU. CoreML chooses to execute the VisionFramework on the GPU instead of the available ANE for the iPhone X, in contrary to the tested MacBook Air, which uses the dedicated accelerator. As such, the frameworks affected by our attack vary from device to device. \\\\
\textbf{Future Work.} Future work in GPU-related vulnerability research can focus on detecting similar vulnerabilities in other implementations of other vendors. Such vulnerabilities of uninitialized interrupt structures have been found in CPUs \cite{borrello2022aepic}. If GPU vendors failed to secure register access, there might also be other architectural structures which are prone to leakage. We further encourage research on leakage of other neural network types, to broaden the scope of the attack.
\section{Related Work}
\label{sec:relatedwork}
In the following, we provide a summary of existing GPU-based attacks and classify them into different categories, and further present existing research on countermeasures. \\\\
\textbf{Architectural side-channel Attacks.} Like CPUs, GPUs have also been shown to suffer from cache side-channel attacks \cite{ahn2021trident, dutta2023spy, wang2023gpu}. Just like their CPU counterparts, GPUs also heavily rely on caches in order to improve system performance.
Researchers have previously been targeting the memory model on NVIDIA GPUs, showing that it is possible to leak data from a victim shader by abusing uninitialized memory \cite{di2013cuda} \cite{maurice2014confidentiality}. These attacks also were able to partially leak some register contents, however, these rely on register spilling to memory. Register spilling is only applied if the executed shader kernel is too big and the number of live variables cannot be hold in the register set anymore. However, the probability for variables to be spilled to memory is quite low, since most GPUs support between 128 and 256 registers per kernel, making register spilling quite unlikely. Just like their CPU counterparts, GPUs are also vulnerable against performance counter attacks \cite{naghibijouybari2018rendered}. Performance counters have also been used in a countermeasure context as workload classification method, in order to detect malicious attempts of cryptomining by malware on a user's system \cite{pott2023overcoming}. To combat advanced attacks on GPUs, there has been academic effort to secure traditional GPUs in loud scenarios \cite{hunt2020telekine, mai2023honeycomb}, however, they often implement their efforts only on AMD GPUs, because of their mostly open-source software stack. \\\\
\textbf{Software-based Frequency Attacks.} Taneja et al. presented a frequency and power side-channel attack targeting GPUs in ARM systems \cite{taneja2023hot}. The researchers showed that the power consuption and the reaction of the GPU's dynamic voltage and frequency scaling (DVFS) mechanisms also relies on the workload and specific data which is processed. This enabled the leaking of pixel values of other applications, as well as profiling the currently visited websites of a user. Further, because the attack relies on frequency measurements, it can even be executed from Javascript code running in the browser. In comparison to our approach, this presented attack has a lower leakage bandwidth, however, it also allows rogue websites to launch the attack. A similar approach has been introduced as a covert channel by González-Gómez et al \cite{gonzalez2023first}. Leveraging the emitted temperature, up to 8.75 bps could be covertly transmitted. \\
An attack leveraging the overclocking capabilities of GPUs has been presented by Sabbagh et al. \cite{sabbagh2020novel}. The overclocking capabilities expose the frequency to set among other tweakable parameters. By scaling the frequency up to a point where some calculations fault and return compromised results, but not crashing the GPU, the researchers were able to break an GPGPU version of AES via differential fault analysis. Qiu et al. leveraged DVFS for attacking deep neural network inference \cite{sun2021lightning}. \\\\
\textbf{Physical Attacks.} Beyond the realm of software, researchers have introduced multiple types of physical attacks targeting GPUs. Researchers presented a power-side-channel on a NVIDIA TESLA GPU by measuring the current consumtion at the ATX power suppy \cite{luo2015side}. Maia et al showed that a phyiscal attacker can extract information about neural networks running on the GPU via a magnetic inductance probe \cite{maia2021can}. More specifically, the researchers were able to extract exact information about the network topology and hyperparameters, but did not leak data processed by these. \\
Research on resistance against physical manipulation attacks on GPUs has been mostly done in context of reliability, often not security. \cite{kastensmidt2016radiation, goncalves2020investigating}. \\\\
\textbf{Countermeasures.} Due to previously discovered vulnerabilities in GPUs, various countermeasures have been proposed which cover a wide range of attack vectors. Researchers aimed to apply the concepts of Trusted Execution Environments (TEEs) also on GPUs, developing multiple GPU TEE implementation in recent years \cite{hunt2020telekine, yudha2022lite, mai2023honeycomb, deng2022strongbox}. Due to their popularity for artificial intelligence, some academic work also focuses on providing confidentiality solely for neural network inference and training via GPUs \cite{asvadishirehjini2022ginn, ng2021goten}. \\
Not only academic researchers, but also industry GPU vendors seem to slowly adapt security concepts known from CPUs. As such, enterprise solutions are now capable of GPU-based virtualization solutions, and NVIDIA announced TEE compatibility for their H100 flagship lineup, which tightly integrates the CPU TEE implementations AMD SEV, Intel TDX and ARM CCA \cite{cch100}. However, published documents does not seem to hint at mitigations for register-based attacks being in place.
\section{Conclusion}
\label{sec:conclusion}
In this paper, we introduced uninitialized register reads as a powerful GPU-based exploit vector, which is capable of leaking vast amounts of processed data. During the attack implementation, multiple obstacles emerged, such as the need to reconstruct the leaked data due to black-box thread group scheduling mechanisms. We presented multiple mechanisms to deal with various types of workloads in our evaluation, thereby giving an adversary the capability reconstruct real-world workloads. We have shown, that such reconstruction mechanisms are able to reconstruct pixel information through stale fragment shader registers, and that such methods can also be applied to complex and large neural networks. Finally, we discussed multiple possible countermeasures against the aforementioned attack to thwart uninitialized register read attacks.

\section*{Responsible Disclosure}
We disclosed our findings to Apple and NVIDIA in July and to Qualcomm in August of 2023. All vendors acknowledged the vulnerabilities found in the paper. At the time of writing, all three vendors were able to reproduce the issue and have stated to be working on fixing the vulnerabilities.

\iffalse
\section*{Data Availability}
Upon acceptance of the paper, we will publish collected data and source code for replicating the vulnerabilities and experiments.
\fi

\bibliographystyle{plain}
\bibliography{references}

%\appendices
%\section{Appendix}
%\subsection{Pixel Reconstruction}

\end{document}